\documentclass[twocolumn,showpacs,preprintnumbers,amsmath,amssymb]{revtex4}
\usepackage{tipa}
\usepackage{amssymb}
\usepackage{txfonts}
\usepackage{hyperref}
\usepackage{graphicx}
\usepackage{epsfig}
\begin{document}

\title{An Effective Two-component Entanglement in Double-well Condensation}
\author{Jing Chen,
 Yanqing Guo, Haijing Cao and Heshan Song\footnote{corresponding author: hssong@dlut.edu.cn}}
\affiliation{Department of Physics, Dalian University of
Technology, Dalian, Liaoning 116023, P.R.China}


\begin{abstract}

We propose a spin-half approximation method for
two-component condensation in double wells to discuss the quantum
entanglement of two components. This approximation is presented to be
valid under stationary tunneling effect for
odd particle number of each component. The evolution of the entanglement is
found to be affected by the particle number both quantitatively and
qualitatively. In detail, the maximal entanglement are shown to be
hyperbolic like with respect to tunneling rate and time. To successively
obtain large and long time sustained entanglement, the particle number
should not be large.

\end{abstract}

\maketitle

\section{INTRODUCTION}
The experimental achievement of atomic Bose-Einstein
condensation(BEC) \cite{1} has opened fascinating possibilities for
studying quantum properties of a macroscopic number of cold
quantum atoms \cite{2} and remains one of the most active research
areas in recent years. Certainly, theoretical attentions are
directed towards the underlying quantum correlation properties of
the condensed atoms \cite{3}. Due to the macroscopic natural
characteristics of BEC, it should be an ideal system for describing
some quantum phenomena related to the
coherence. Correspondingly, although the Bose-Einstein condensate is
well represented by mean-field theory, it has many aspects that can
be represented in a quantum picture containing some proper
description of correlations. The essential ultimate physics can be
realized with the study of a simpler double-well BEC in the
well-known quantum two-mode approximation.

As is well known, entanglement has come to be regarded as a
physical resource which can be utilized to perform numerous tasks
in quantum information processing. Also, apart from the fundamental
physical interest in entanglement, the whole field of quantum
computing and quantum information is based upon the ability to
create and control entangled states \cite{4}. In recent times, the
study of the entanglement characteristics of various condensed
matter system \cite{5} is focused on a pair of tunnel-coupled
Bose-Einstein condensates (BEC's). In the simplest model,
bosons are restricted to occupy one of two modes, each of
which is in a BEC \cite{6}.

A dynamical scheme of engineering many-particle entanglement in
BEC has been proposed by several authors, such as, Khan W. Mahmud \emph{et al} \cite{7}.
They introduced a quantum phase-space dynamics to
generate tunable entangled number states using the Husimi
projection method \cite{8}. However, the notion of entanglement
in macroscopic ensembles allows to investigate the boundary
between quantum physics and classical physics and, possibly, could
also give some insight into the measurement process \cite{9,10,11}.
In this paper we will investigate in detail a
scheme that measures entanglement of $N$ two-component particles
trapped in a double-well under a kind of reasonable
approximate assumption.This approximation is presented to be
valid under stationary tunneling effect for
odd particle number of each component.

\section{THE MODEL}

Our system consists of a double-well and two component condensate
trapped in it as depicted in FIG. 1.
\begin{figure}
\epsfig{file=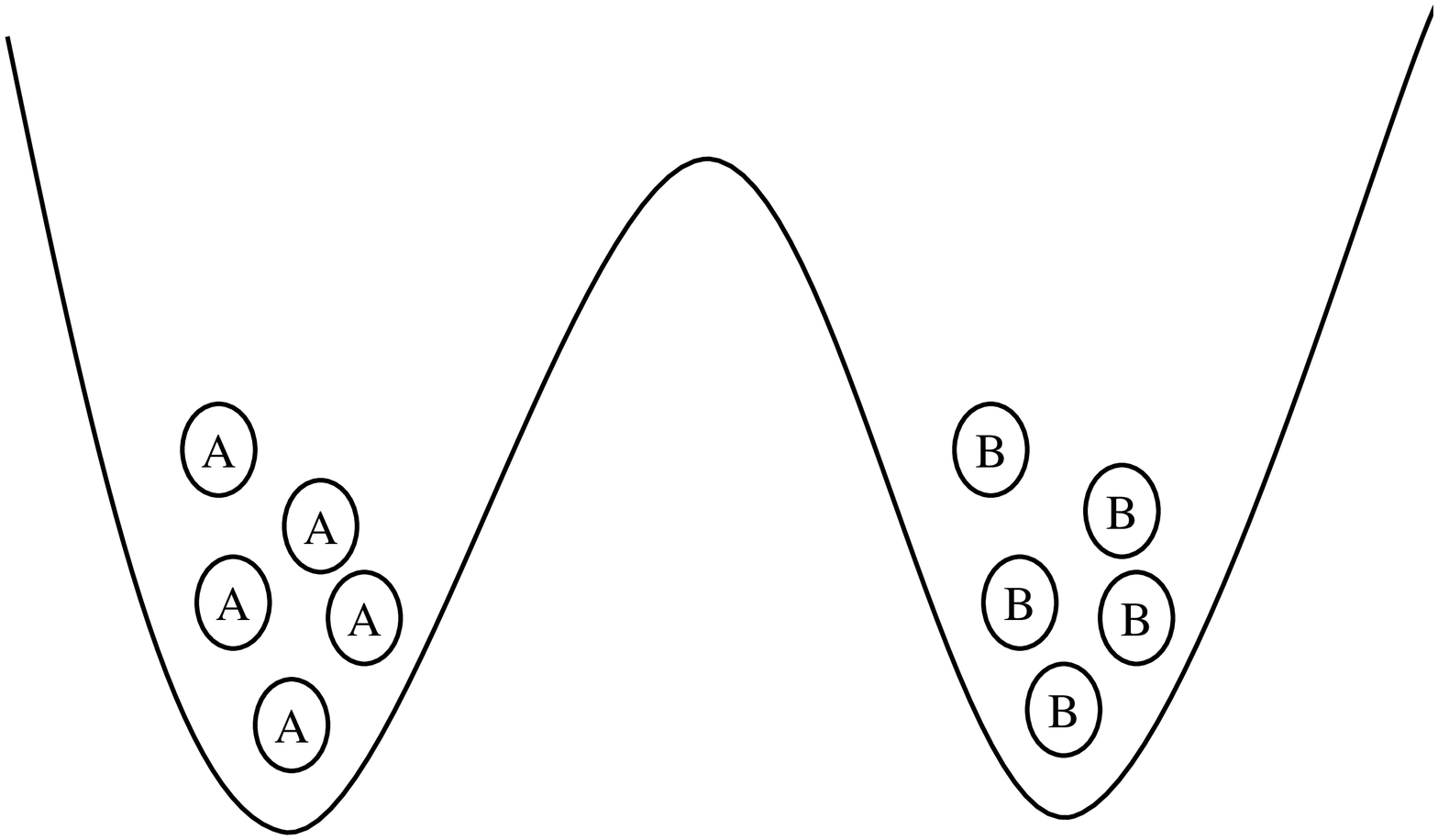, width=7cm, height=5cm}
\caption{{\protect\footnotesize {A sketch of the double-well system
with two-component condensate trapped in. Initially, the atoms of the
component A are trapped in the left side and the atoms of the component B
are trapped in the right side.}}}
\end{figure}
Initially, the atoms of component A and B are separately located in
left and right potential wells, respectively, \cite{12}. The many-body Hamiltonian
for a system of $N$ weakly interacting bosons
in an external potential $V(r)$, in second quantization, is given by \cite{7}
\begin{eqnarray}
\hat{H}&=&\int{d\boldmath{r}\hat{\psi^{\dag
}}(\boldmath{r})\bigg[-\frac{\hbar^{2}}{2m}\nabla^{2}+V(\boldmath{r})\bigg]
\hat{\psi}(\boldmath{r})}\nonumber\\
&&+\frac{g}{2}\int{d\boldmath{r}\hat{\psi^{\dag
}}(\boldmath{r})\hat{\psi^{\dag }}(\boldmath{r})
\hat{\psi}(\boldmath{r})\hat{\psi}(\boldmath{r})},
\end{eqnarray}
where $\hat{\psi}(\boldmath{r})$ and $\hat{\psi^{\dag
}}(\boldmath{r})$ are the boson annihilation and creation field
operators, $m$ is the particle mass and $g=\frac{4\pi
a_s\hbar^{2}}{m}$, with $a_s$ the s-wave scattering length. In
studies of double-well BEC or two-component condensates, the
low-energy many-body Hamiltonian in Eq. (1) can be simplified in the
well-known two-mode approximation \cite{13,14}, which has been used widely
in studying the double-well condensate. Under this approximation, the system is
modeled by the Hamiltonian($\hbar=1$) \cite{12},
\begin{eqnarray}
\hat{H}&=&\frac{\Omega_{A}}{2}(a_{L}^{+}a_{R}+a_{R}^{+}a_{L})+\frac{\Omega_{B}}{2}(b_{L}^{+}b_{R}+b_{R}^{+}b_{L})
+\kappa(a_{L}^{+}a_{L}b_{L}^{+}b_{L}\nonumber\\
&&+a_{R}^{+}a_{R}b_{R}^{+}b_{R})+\frac{\kappa_{A}}{2}[(a_{L}^{+}a_{L})^{2}+(a_{R}^{+}a_{R})^{2}]\nonumber\\
&&+\frac{\kappa_{B}}{2}[(b_{L}^{+}b_{L})^{2}+(b_{R}^{+}b_{R})^{2}],
\end{eqnarray}
where the subscripts L and R denote respectively the localized modes in the left and right potential wells.
Because of two modes available for each component, there are four operators $a_{j}^{+}$ and $b_{j}^{+}$
($j=L,R$) that denote the creation operators of the components A and B in two wells in the model.
The parameters $\Omega_{i}$, $\kappa_{i}$ with $i=A,B$ and
$\kappa$ describe the tunneling rate, self-interaction strength of the component A (B) and
the interspecies interaction strength.
Especially, $\Omega_{i}$ depends linearly on particle number $N_{i}$, so that,
$\Omega_{i}=\epsilon_{LR}+gT_{1}(N_{i}-1)$,
where $\epsilon_{LR}$, $T_{1}$ are the parameters after second quantization and $N_{i}$ is particle
number of component $i$ \cite{7}.
For simplicity, we set $\Omega_{A}=\Omega_{B}=\Omega$ in following sections.

\section{CLASSIC ANALYSIS}

Due to the macroscopic nature of its wave function, BEC should be an ideal system for the generation of
macroscopic quantum superposition states (Schr\"{o}dinger cat states) \cite{7}. Such a system could be,
on one hand, analyzed in a classical method from a large particle number point of view and on the other
hand studied in a quantum method because of the quantum superposition character.

The classical Hamiltonian that describes the mean-field dynamics of BEC in a double well has been studied in several
papers \cite{15,16}. In a mean-field assumption for the two-mode double well BEC in case of large enough $N$, the operators $\hat{a}_{j}$
can be replaced by a \emph{c}-number $\sqrt{n_{j}} e^{i\theta_{j}}$, where $j=L,R$. With this assumption and defining population difference (POD)
$n=\frac{(n_{L}-n_{R})}{2}$, $\theta=\theta_{L}-\theta_{R}$, we analogously introduce \emph{c}-numbers both for component A and
B, that is, $\hat{a}_{j}\rightarrow\sqrt{n_{Aj}} e^{i\theta_{Aj}}$, $\hat{b}_{j}\rightarrow\sqrt{n_{Bj}} e^{i\theta_{Bj}}$,
\begin{eqnarray*}
&&n_{AL}+n_{AR}=N_{A},\\
&&n_{BL}+n_{BR}=N_{B},\\
&&n_{A}=\frac{n_{AL}-n_{AR}}{2},\\
&&n_{B}=\frac{n_{BL}-n_{BR}}{2},
\end{eqnarray*}
the classical Hamiltonian is then given by
\begin{eqnarray}
H_{cl}&=&\frac{\Omega}{2}\bigg[\sqrt{N_{A}^{2}-(2n_{A})^{2}}\cos\theta_{A}+\sqrt{N_{B}^{2}-(2n_{B})^{2}}\cos\theta_{B}\bigg]\nonumber\\
&&+\frac{\kappa}{2}N_{A}N_{B}+\frac{\kappa_{A}}{4}N_{A}^{2}+\frac{\kappa_{B}}{4}N_{B}^{2}\nonumber\\
&&+\kappa_{A}n_{A}^{2}+\kappa_{B}n_{B}^{2}+2\kappa n_{A}n_{B}
\end{eqnarray}
and the equations of motion are
\begin{eqnarray}
\dot{n}_{A} &=& \frac{\Omega}{2}\sqrt{N_{A}^{2}-(2n_{A})^{2}}\sin\theta_{A},\\
\dot{n}_{B} &=& \frac{\Omega}{2}\sqrt{N_{B}^{2}-(2n_{B})^{2}}\sin\theta_{B},\\
\dot{\theta_{A}}&=&-\frac{2\Omega n_{A}\cos\theta_{A}}{\sqrt{N_{A}^{2}-(2n_{A})^{2}}}+2\kappa_{A}n_{A}+2\kappa n_{B},\\
\dot{\theta_{B}}&=&-\frac{2\Omega n_{B}\cos\theta_{B}}{\sqrt{N_{B}^{2}-(2n_{B})^{2}}}+2\kappa_{B}n_{B}+2\kappa n_{A}.
\end{eqnarray}
We plot
the amount of the POD for component A and B ($n_{A},n_{B}$) as a function of the
tunneling rate $\Omega$ and time $t$ in FIG. 2 and FIG. 3.
\begin{figure}
\epsfig{file=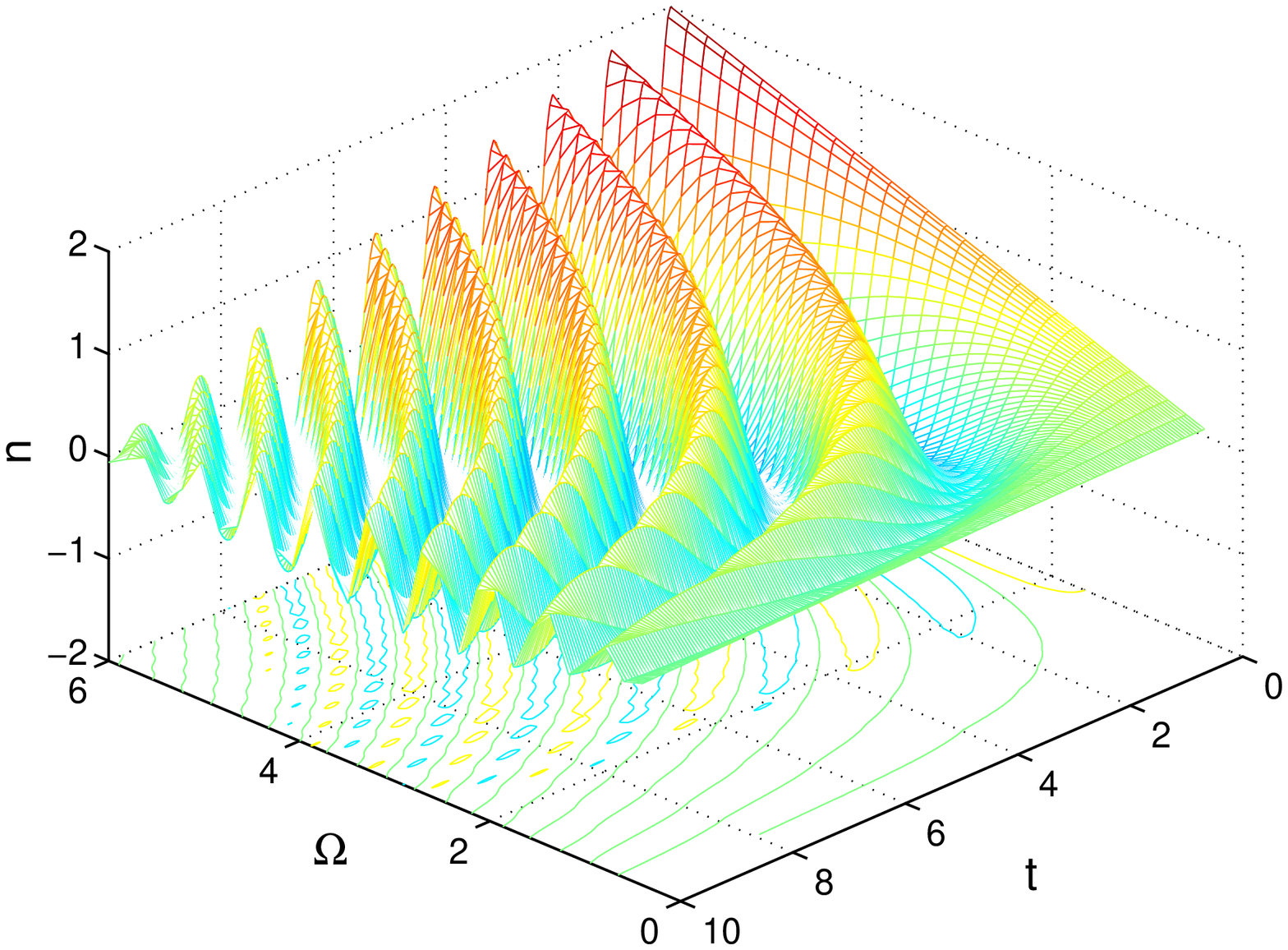, width=6cm, height=6cm}
\caption{{\protect\footnotesize {Shown is the population difference for component A with
parameter values $\kappa=20$, $\kappa_{A}=20$, $\kappa_{B}=20$.}}}
\end{figure}

\begin{figure}
\epsfig{file=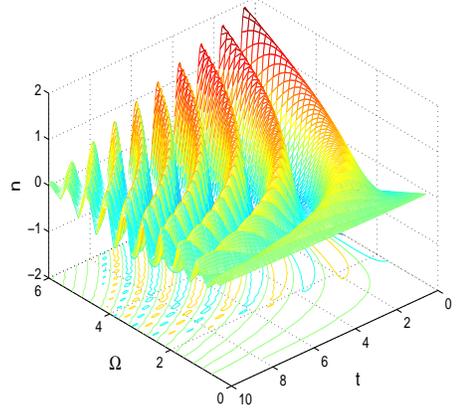, width=6cm, height=6cm}
\caption{{\protect\footnotesize {Shown is the population difference for component B for
parameter values $\kappa=20$, $\kappa_{A}=20$, $\kappa_{B}=20$.}}}
\end{figure}

FIG. 2 shows the POD for component A with
parameter values $\kappa=20$, $\kappa_{A}=20$, $\kappa_{B}=20$.
The POD between two wells of each component presents a behavior of local maximum and minimum.
The period is determined by the particle number of the component. The larger the particle number, the smaller the period.
It should be stressed that the peak of the POD of component A between two wells is
decreased gradually.
In FIG. 3, there have the same conditions for component B but for
the negative initial value of POD.

FIG. 2 suggests that the stationary state of population would be in particle number equilibrium. Similar result
has been presented in Ref. [12], where the expectation value of the defined inversion
operator $\langle a_{L}^{+}a_{L}-a_{R}^{+}a_{R} \rangle$ was found to tend to zero.
Our solution is based on a continuous analysis of Equ. [4-7]. While, in physical system scenario,
the population number must be discrete
since there are $N+1$ orthogonal Fork states of each component. That is, for even component particle number $N$,
the stationary population
in each well would be $\frac{N}{2}$. While for odd $N$, the stationary POD would be $\pm\frac{1}{2}$.
Whatever be the initial population number of components in each well, a minimal POD-0 or 1 can be
obtained due to the tunneling effect and repulsive interaction.

In the following section, we shall see how to generate and measure a mode entangled state under such
an approximate situation.

\section{GENERATING AND MEASURING OF ENTANGLEMENT UNDER SPIN-HALF APPROXIMATION ASSUMPTION}

The quantum phase-space model presented in Ref. [7] points to a way that an entangled state can
be generated with a single-component BEC in double well. The authors showed the generation of
tunable entangled states in phase space using Husimi distribution function. We now introduce
another method to prepare entangled states. Schwinger has developed an entire angular-
momentum algebra in terms of two sets (up and down) of creation
and annihilation operators for uncorrelated harmonic oscillator constructing the angular-momentum
operators as \cite{17}
\begin{eqnarray}
&&J_{+}=J_{x}+iJ_{y}=a_{+}^{+}a_{-}\nonumber,\\
&&J_{-}=J_{x}-iJ_{y}=a_{-}^{+}a_{+}\nonumber,\\
&&J_{z}=\frac{1}{2}(a_{+}^{+}a_{+}-a_{-}^{+}a_{-}).
\end{eqnarray}

This construction then satisfies the standard angular-momentum commutation
relation $\big[J_{l}, J_{m}\big]=i\epsilon_{lmn}J_{n}$.
Now, we regard the double-well as a two-state system and
similarly introduce an analogous-angular momentum algebra to express the Hamiltonian in Eq. [2].
For the two-component condensate trapped in a double well,we defining
\begin{eqnarray}
&&S_{i}^{+}=\Lambda_{L}^{+}\Lambda_{R},\nonumber\\
&&S_{i}^{-}=\Lambda_{L}\Lambda_{R}^{+},\nonumber\\
&&S_{i}^{z}=\frac{1}{2}(\Lambda_{L}^{+}\Lambda_{L}-\Lambda_{R}^{+}\Lambda_{R}),
\end{eqnarray}
where $i$=A, B, $\Lambda=a, b$ (for components A and B respectively). $S_{\Lambda}^{+}$, $S_{\Lambda}^{-}$
satisfy the angular-momentum commutation relations:
$\Big[S_{\Lambda}^{+}, S_{\Lambda}^{-}\Big]=2S_{\Lambda}^{z}$.
With these descriptions, we, in fact, consider the left and right wells as two states ($|L\rangle$ and $|R\rangle$) of
the particles. As for $N$-particle system, the collective spin is $S_{z}=-\frac{N}{2}\cdots \frac{N}{2}$.
So, there would be $N_{i}$ ``excited states'' for
component $i$ with
particle number $N_{i}$ as illustrated in FIG. 4.
\begin{figure}
\epsfig{file=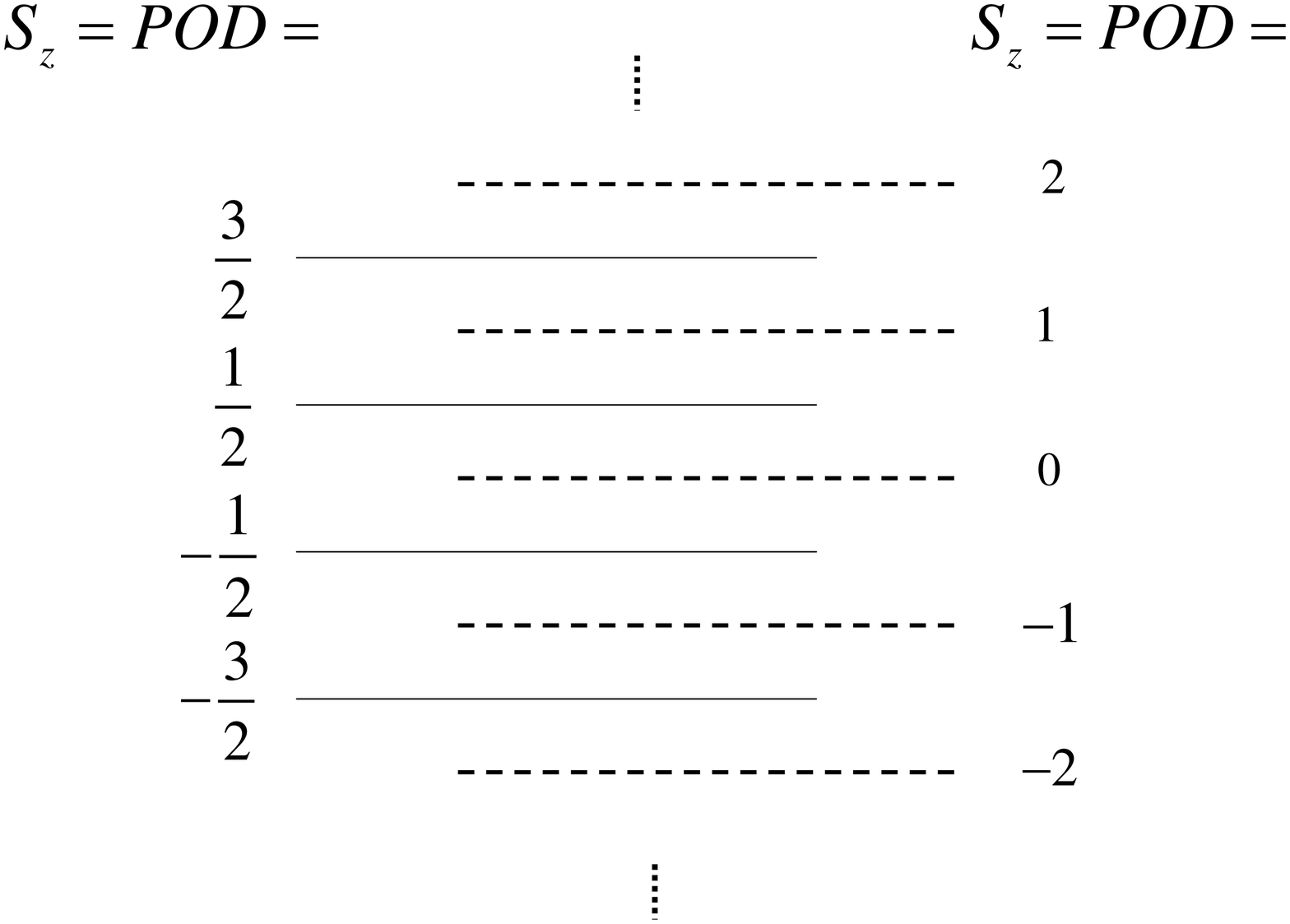, width=5cm, height=5cm}
\caption{{\protect\footnotesize {Shown is the distribution of
pseudo-spin $S_{z}$ with the respect to POD. Solid line for odd particle number, dotted line for even particle number.}}}.
\end{figure}
Now, we assume the initial particle number of each component is odd. The stationary POD,
from the analysis in the last section, must be 1. So, the collective spin of each component should be $\pm\frac{1}{2}$.
This assumption suggests us to eliminate adiabatically ``high energy'' states of each component,
thus each component can be treated as a pseudo-particle with expectation value $\pm\frac{1}{2}$ of spin.

Then, the raising and lowering operators of this pseudo-particle $S^{+}$ $(S^{-})$ can be expressed in a $2\times2$
dimensional Hilbert space by
\begin{eqnarray}
&&S_{A}^{+}\rightarrow|L\rangle_{A}\langle R|, ~ ~S_{A}^{-}\rightarrow|R\rangle_{A}\langle L|\nonumber,\\
&&S_{B}^{+}\rightarrow|L\rangle_{B}\langle R|, ~ ~S_{B}^{-}\rightarrow|R\rangle_{B}\langle L|.
\end{eqnarray}

After the operators $\hat{a}_{i}^{+}(\hat{a}_{i})$ replaced by the $S_{j}^{+}(S_{j}^{-})$, where $j=A,B, i=L,R$,
the global Hamiltonian in Equ. [2] can be spanned on the basis of
 $|L\rangle_{A}\otimes|L\rangle_{B}$, $|L\rangle_{A}\otimes|R\rangle_{B}$, $|R\rangle_{A}\otimes|L\rangle_{B}$, $|R\rangle_{A}\otimes|R\rangle_{B}$.
We obtain the effective Hamiltonian of our system as
\begin{eqnarray}
H=\left(\begin{array}{cccc}
K_{1} & K & K & 0 \\
K & K_{2} & 0 & K \\
K & 0 & K_{2} & K \\
0 & K & K & K_{1}
\end{array}\right)
\end{eqnarray}
where $K=\frac{\Omega}{2}$, $K_{1}=\kappa+\frac{\kappa_{A}}{2}+\frac{\kappa_{B}}{2}$, $K_{2}=\frac{\kappa_{A}+\kappa_{B}}{2}$.
The eigenvalues and eigenvectors of the Hamiltonian can be found explicitly,
\begin{eqnarray}
&&E_{1}=K_{2}, ~\phi_{1}=-\frac{1}{\sqrt{2}}\big(|L\rangle_{A}|R\rangle_{B}-|R\rangle_{A}|L\rangle_{B}\big);\nonumber\\
&&E_{2,3}=\frac{K_{2}+K_{1}\pm\Theta}{2},\nonumber\\
&&\phi_{2,3}=\xi_{2,3}\bigg[|L\rangle_{A}|L\rangle_{B}\mp\frac{-K_{2}+K_{1}-\Theta}{4K}\nonumber\\
&&\big(|L\rangle_{A}|R\rangle_{B}+|R\rangle_{A}|L\rangle_{B}\big)+|R\rangle_{A}|R\rangle_{B}\bigg];\nonumber\\
&&E_{4}=K_{1},~ \phi_{4}=\frac{1}{\sqrt{2}}\big(|L\rangle_{A}|L\rangle_{B}-|R\rangle_{A}|R\rangle_{B}\big).
\end{eqnarray}
where $\xi_{2}$, $\xi_{3}$ are normalized factors and $\Theta=\sqrt{K_{2}^{2}-2K_{1}K_{2}+K_{1}^{2}+16K^{2}}$.
We assume that, initially, the POD of component A is $\frac{1}{2}$, and the POD of component B is $-\frac{1}{2}$,
so that, the initial state of system is
\begin{eqnarray}
|\psi(0)\rangle=|L\rangle_{A}\otimes|R\rangle_{B}.
\end{eqnarray}

\begin{figure}
\epsfig{file=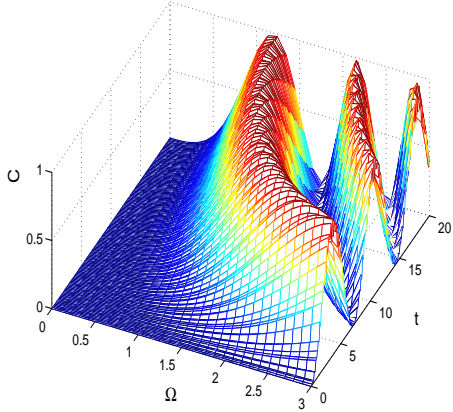, width=6cm, height=6cm}
\caption{{\protect\footnotesize {Shown is the degree of entanglement as a
function of the tunneling rate $\Omega$ and time $t$
under the two-mode approximate Hamiltonian for
parameter values $\kappa=20$, $\kappa_{A}=20$, $\kappa_{B}=20$.}}}.
\end{figure}

\begin{figure}
\epsfig{file=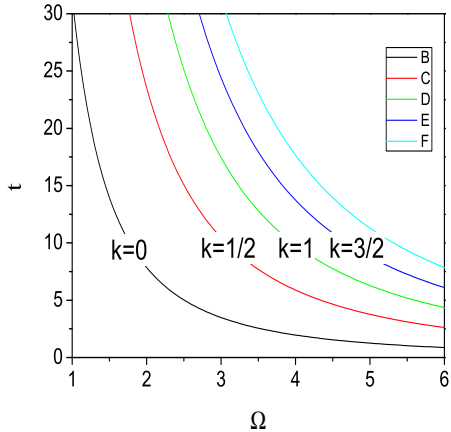, width=7cm, height=7cm}
\caption{{\protect\footnotesize {Shown is the maximal entanglement for different $k$ as a
function of the tunneling rate $\Omega$ and time $t$
under the two-mode approximate Hamiltonian for
parameter values $\kappa=20$, $\kappa_{A}=20$, $\kappa_{B}=20$.}}}
\end{figure}

Wootters Concurrence has been widely used in measuring the entanglement of bipartite two-state system,
which is defined as \cite{18},
\begin{eqnarray}
C(\rho)=\max\big\{0, \lambda_{1}-\lambda_{2}-\lambda_{3}-\lambda_{4}\big\},
\end{eqnarray}
where $\lambda_{i}$ are the square roots of he eigenvalues of the non-Hermitian matrix
$\rho\widetilde{\rho}$ with $\widetilde{\rho}=(\sigma_{y}\otimes\sigma_{y})\rho^{*}(\sigma_{y}\otimes\sigma_{y})$
in decreasing order. Wootters Concurrence gives an explicit expression for the entanglement of formation,
which quantifies the resources needed to create entangled state.

Then the behavior of entanglement in this system described by
spin-half assumption Hamiltonian is illustrated in FIG. 5 and FIG. 6.
In this situation when there is no decoherence, entangled states can be generated because of the overlap
of the wave packet of each component in two wells. We note that the entanglement
presents a periodic distribution for any value of $t$ falling in the region, that is,
the behavior of the amount of entanglement between the two modes
is non-monotonic, the maximum entanglement can be achieved if we control the tunneling time at about
$t=\frac{2\kappa}{\Omega^{2}}\big(\frac{\pi}{4}+\frac{k\pi}{2}\big)$.
The corresponding maximal entanglement for different $k$ described by FIG. 6 is shown to be
hyperbolic like with respect to tunneling rate and time.
The half width of the peak of entanglement $\tau$ can be defined as a parameter to
represent the quantity and the quality of the maximal entanglement.
As an example, we show the entanglement when $\Omega=1$ in FIG. 6.
At the peak position of the entanglement, $t\approx\frac{2\kappa}{\Omega^{2}}\big(\frac{\pi}{4}+\frac{k\pi}{2}\big)$,
the half width is $\tau\approx31.369$. While for $\Omega=1.5$, the half
width is $\tau\approx13.899$, which suggests that the larger the $\Omega$, the smaller the half
width of the peak.
Then, we can easily obtain
large and long time sustained entanglement successively for $1\leq\Omega\leq1.5$,
which presents a broad banding distribution.

\begin{figure}
\epsfig{file=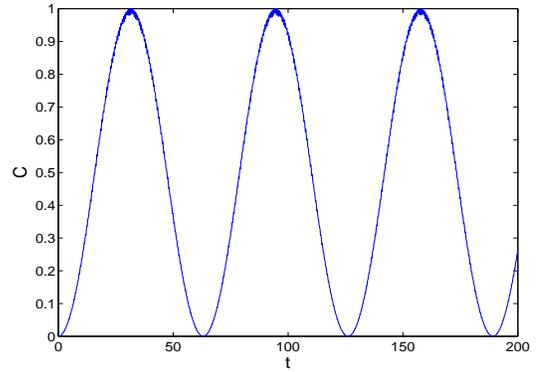, width=7cm, height=5cm}
\caption{{\protect\footnotesize {Shown is the degree of entanglement as a function of time t
under the two-mode approximate Hamiltonian for
parameter values $\Omega=1$, $\kappa=20$, $\kappa_{A}=20$, $\kappa_{B}=20.$}}}.
\end{figure}

\section{CONCLUDING REMARKS}
The creation of many particle entangled states in macroscopic
systems is one of the major goals in the studies on fundamental
aspects of quantum theory. In this paper,
the studies we have performed show a spin-half approximation method for
two-component condensation in double wells to discuss the quantum
entanglement of two components,
which is presented to be
valid under stationary tunneling effect for
odd particle number of each component. The evolution of the entanglement is
found to be affected by the particle number both quantitatively and
qualitatively. In detail, the maximal entanglement are shown to be
hyperbolic like with respect to tunneling rate and time. To successively
obtain large and long time sustained entanglement, the particle number
should not be large.
Thus we can transform an exact
many-body problem which is difficult even for single-component condensate to a bipartite two-state problem,
similarly, we could measure the entanglement simply and conveniently. There also refer to the comparison of classical and
quantum dynamics, great progress has been made in this subject \cite{7}, which is interesting and sensitive recently
and is good for further studies.
Our entanglement states are specified with the particle spatial location in left or right well rather than internal energy levels,
this may be operational in physical applications such as quantum entangled particles distribution and quantum measurement.
Moreover, the method developed here may find applications in the studies of entanglement of other
BEC.

\section{Acknowledgments}

This work was supported by NSF of China, under grant No. 60472017.

\end{document}